**Presolar Diamond in Meteorites**

*Sachiko Amari*[A,B]

[A]Laboratory for Space Sciences and the Physics Department, Washington University, St. Louis, MO 63130, USA

[B]E-mail: sa@wuphys.wustl.edu

**Abstract.** Presolar diamond, the carrier of the isotopically anomalous Xe component Xe-HL, was the first mineral type of presolar dust that was isolated from meteorites. The excesses in the *l*ight, p-process only isotopes $^{124}$Xe and $^{126}$Xe, and in the *h*eavy, r-process only isotopes $^{134}$Xe and $^{136}$Xe relative to the solar ratios indicate that Xe-HL was produced in supernovae: they are the only stellar source where these two processes are believed to take place. Although these processes occur in supernovae, their physical conditions and timeframes are completely different. Yet the excesses are always correlated in diamond separates from meteorites. Furthermore, the p-process $^{124}$Xe/$^{126}$Xe inferred from Xe-L and the r-process $^{134}$Xe/$^{136}$Xe from Xe-H do not agree with the p-process and r-process ratios derived from the solar system abundance, and the inferred p-process ratio does not agree with those predicted from stellar models. The "rapid separation scenario", where the separation of Xe and its radiogenic precursors Te and I takes place at the very early stage (7900 sec after the end of the r-process), has been proposed to explain Xe-H. Alternatively, mixing of 20% of material that experienced neutron burst and 80% of solar material can reproduce the pattern of Xe-H, although Xe-L is not accounted for with this scenario.

# 1. Introduction

Our solar system was once thought an isotopically homogenous system. Although the molecular cloud, from which the solar system formed, contained dust and gas expelled from stars, it had been believed that dust was evaporated and all the material in the cloud had been well mixed during the formation of the solar system, and the subsequent events erased any remaining trace of heterogeneity, if any. However, in the late 1960's, when Black and Pepin (Black & Pepin 1969) analyzed Ne in primitive meteorites, they found that Ne with low $^{20}$Ne/$^{22}$Ne (down to 3.4) was released between 900 and 1100°C fractions that could not be explained by the Ne components commonly observed in meteorites. They concluded that a new component with a low $^{20}$Ne/$^{22}$Ne ratio was present in the meteorites and named this new component Ne-E. Subsequent studies showed that there are two kinds of Ne-E, Ne-E(H) and Ne-E(L). Ne-E(H) is released in *h*igher temperature steps (1200 – 1400°C) and is concentrated in *h*igher-density mineral separates (3 – 3.5 g/cm$^3$), whereas Ne-E(L) is released in *l*ower temperature steps (500 – 700°C) and is concentrated in *l*ower density separates (< 2.3 g/cm$^3$) (Jungck 1982).

For isotopic anomalies in heavy noble gases, Lewis et al. (1975) observed $^{124}$Xe/$^{130}$Xe and $^{136}$Xe/$^{130}$Xe ratios increased almost two fold compared with normal Xe in a chemically treated residue obtained from the Allende meteorite. This highly anomalous component was named Xe-HL because both the *l*ight isotopes $^{124}$Xe and $^{126}$Xe, and the *h*eavy isotopes $^{134}$Xe and $^{136}$Xe were enriched (Figure 1). Other isotopically anomalous noble gas components include Kr-S (s-process Kr) (Srinivasan & Anders 1978; Alaerts et al. 1980) and Xe-S (s-process Xe) (Srinivasan & Anders 1978).

The fact that these isotopically anomalous components were well hidden in a vast amount of close-to-normal components in bulk meteorites suggests that minerals with these components were very rare in meteorites. The pursuit of the carriers of these noble gas components was carried out by Edward Anders and Roy S. Lewis and their colleagues at the University of Chicago [for historical accounts of the discovery of the anomalous noble gas components and their carriers, see Anders 1988]. Their procedure was to chemically and physically remove other minerals. During the process the anomalous noble gases were used as tracers to search for the carrier minerals. Finally, Lewis et al. (1987) isolated the carrier of Xe-HL, diamond, in 1987. This discovery was followed by isolation and identification of SiC (Bernatowicz et al. 1987; Tang & Anders 1988), the carrier of Kr-S, Xe-S and Ne-E(H), and graphite (Amari et al. 1990), the carrier of Ne-E(L). Those minerals are among what we call presolar grains. They are defined as dust that formed in stellar outflow or stellar ejecta and were later incorporated into meteorites, surviving the events during the solar system formation and metamorphism in meteorite parent bodies. Other presolar grains identified in meteorites to date include oxides, $Si_3N_4$, refractory carbides as sub-grains in SiC and graphite, and silicates. Studies of presolar grains have yielded a wealth of information about nucleosysnthesis in stars, grain formation, Galactic chemical evolution and mixing in stellar ejecta (e.g., Lodders & Amari 2005; Zinner 2007).

Presolar diamond was the first mineral type to be isolated from meteorites and it is relatively abundant (~1000 ppm in primitive meteorites) compared with other carbonaceous presolar dust (a few ppm) (Huss & Lewis 1995). However, it remains one of the least understood presolar grains for several reasons. First, because of an extremely small grain size (3 nm in diameter) (Daulton et al. 1996), isotopic analyses of individual grains are not possible and only bulk analyses have been performed. This makes it hard to correlate isotopic features of different elements because we do not know whether isotopic anomalies of different elements originate from the same suite of diamond particles. Second, due to low trace element abundances in diamond (Lewis et al. 1991), isotopic analyses of trace elements are challenging because the analyses are compromised even by a very small amount of impurities. In this paper, we will briefly summarize previous studies on presolar diamond and review outstanding problems, focusing mainly on nucleosynthetic aspects of the problems.

**2. Noble gas components and C and N isotopic ratios in presolar diamond and its origin**

Diamond is highly resistant to chemicals, is thus extracted from meteorites by removing almost everything else, including dissolving silicates with HF and destroying organic matter with oxidants. Although Xe-HL is the most distinct Xe component in diamond from meteorites, detailed studies showed that diamond contains other close-to-normal components. Huss and Lewis (1994a) analyzed noble gases in diamond separates from 14 primitive chondrites in 7 different groups (CI, LL, L, H, EH, CV, and CO) with stepped pyrolysis and identified three noble-gas components which consist of five noble gases. P3 is released in low temperatures (200-900°C) and has a planetary elemental abundance pattern (=enriched in the heavy noble gases compared with solar) and "normal" isotopic ratios. HL is released between 1100 and 1600°C. P6 is released at a slightly higher temperature than HL with roughly "normal" isotopic compositions and "planetary" elemental ratios. The presence of the three noble-gas components strongly indicates that diamonds originated from more than one sources.

Adding a diamond separate from Murray (CM2) to their database, Huss and Lewis (1994b) examined abundances of these noble gas components. P3 gases are most abundant in diamond separates from CI and CM, having roughly the same P3 abundances. Separates from other compositional groups have lower P3 abundances. Among the same group, the P3 abundances decrease with increasing petrologic subtype, indicating that P3 was lost during thermal metamorphism. In contrast, abundances of HL and P6 are remarkably constant through different metamorphic grades and compositional groups, varying only a factor of 1.5 – 1.8 and 2 – 3, respectively. Obviously HL and P6 are more resistant to thermal processing. They proposed a model where all meteorites had contained a single diamond mixture (that seems to be preserved in CI and CM2 chondrites) and various degrees of thermal processing yielded different P3 abundances.

In contrast to the isotopically distinct component Xe-HL, the isotopic ratio of the major element C as well as that of N are bewilderingly close to normal. Russell et al. (1996) analyzed C and N isotopic ratios of diamond-rich separates (containing 15 – 50% diamond) from carbonaceous, ordinary and enstatite chondrites by stepped combustion. $\delta^{13}C$ values varied between –32 to –38 ‰, while $\delta^{15}N$ values were determined to be –348 ± 7 ‰, surprisingly close to the solar ratios despite the highly anomalous Xe-HL (δ values are defined as deviations from the normal ratios by thousand: $\delta^{13}C \equiv [(^{13}C/^{12}C)_{sample}/ (^{13}C/^{12}C)_{standard} – 1] \times 1000$, and $\delta^{15}N \equiv [(^{15}N/^{14}N)_{sample}/ (^{15}N/^{14}N)_{standard} – 1] \times 1000$). They argued that the complexity of the C release pattern and C/N ratio imply the presence of more than one component, and that the components are present in varying proportions in different meteorite groups. The presence of the close-to-normal Xe components and C isotopic ratios indicate that there may be more than one source of presolar diamond.

The presence of Xe-HL indicates that Xe-HL is closely associated with supernovae because they are the only source where both p- and r-processes take place. However, the places where all the diamond particles formed are not well determined although supernovae are one of the likely sources of diamond formation. Another possible source is the interstellar medium (ISM). A 3.4 – 3.5 μm infrared absorption feature in Elias 1, HD 97048 and dense clouds have been attributed to diamond-like material because the surface of pure diamond crystals have a high reactivity with H, giving rise to CH stretching vibrations in the spectral range (e.g., Oliver et al., 2007). Investigating laboratory spectra and astrophysical spectra, Oliver et al. (2007) concluded that their results support the 3.43 and 3.53 μm emission features are due to diamondoids of a few nm in size and the 3.47 μm absorption feature can be attributed to smaller diamondoid molecules. We note, however, it remains to be seen whether part of presolar diamond in meteorites originated from the ISM, and not in the stellar outflow/ejecta.

## 3. Xe-HL

### 3.1. Association of Xe-H and Xe-L

As described in the previous section, the presence of the few noble gas components indicates that diamond had experienced a complex history. Of the Xe components in diamond, Xe-HL is most anomalous, having the excess in the light, p-process only isotopes, 124 and 126 (referred as Xe-L), and that in the heavy r-process only isotopes, 134 and 136 (Xe-H) (Figure 1). Since supernovae are the only place where these two processes take place, it has been believed that Xe-HL and diamond were produced in supernovae.

p-Process takes place in the C-burning zone (or O/Ne zone, whose name indicates most abundant elements, coined by Meyer et al. (1995)) (Rayet et al. 1995). It is photo-disintegration processes [(γ,n), (γ,p) and (γ,α)] of pre-existing neutron-rich nuclei under high temperature (T > ~ $2 \times 10^9$ K).

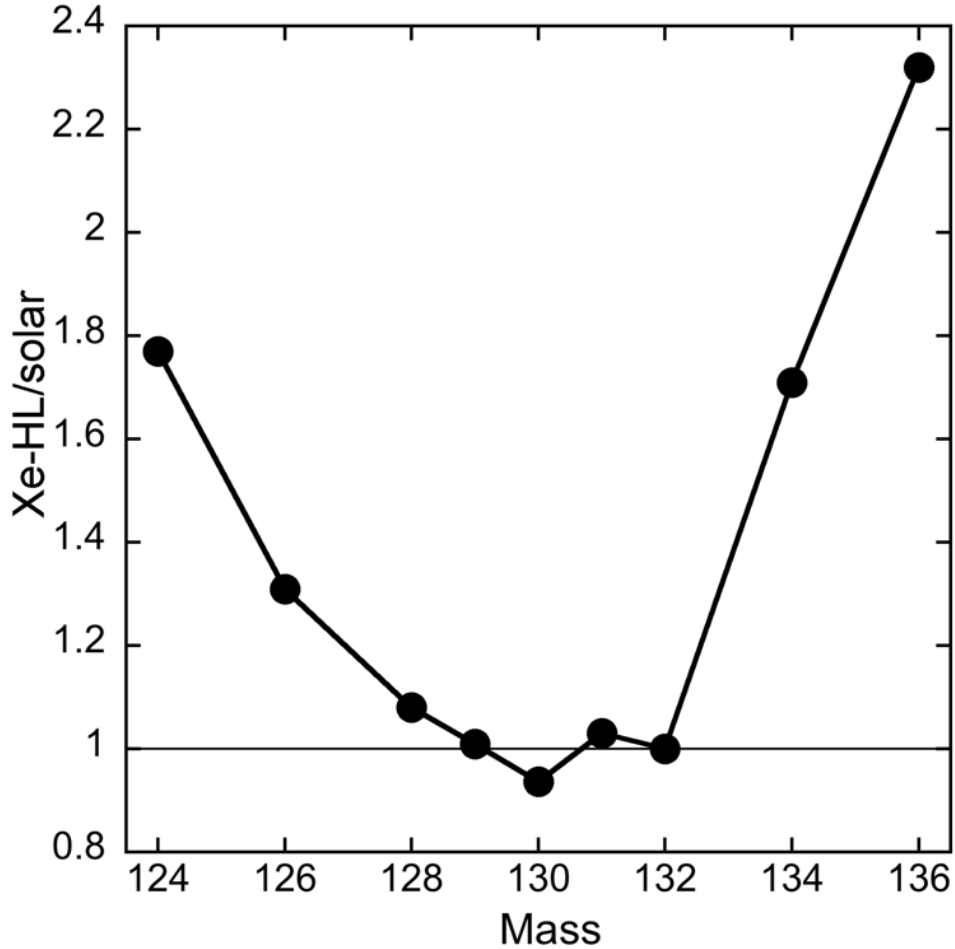

Figure 1. Xe-HL (Huss & Lewis, 1994a). Xenon isotopes are normalized by $^{132}$Xe and by the solar ratios. Xe-HL shows pronounced excesses in the light isotopes $^{124}$Xe and $^{126}$Xe, and in the heavy isotopes $^{134}$Xe and $^{136}$Xe. The excess in $^{124}$Xe and $^{126}$Xe is often referred as Xe-L, that in $^{134}$Xe and $^{136}$Xe as Xe-H.

On the other hand, the place where the r-process takes place has not been identified. Several r-process sources have been proposed. Stars that experience prompt explosion are one of such sources (Wheeler, Cowan, & Hillebrandt 1998; Wanajo et al. 2003). Wanajo et al. (2003) simulated energetic prompt explosion (prior to the neutrino heating) on O-Ne-Mg cores in stars with lower-masses (8-10M$_{sun}$), which led to robust production of r-process nuclei whose relative abundances agree with the solar r-process pattern, in particular, in nuclei with A > 130. However, the production of the r-process matter is about 2 orders of magnitude higher than what is expected from the Galactic chemical evolution. To resolve this, they proposed that most of the r-process material falls back to the proto-neutron star. It remains to be seen whether the prompt explosion, caused by an artificial enhancement of the shock-heating energy, indeed takes place.

Another candidate for the r-process site, a promising one, is neutron-driven winds from nascent neutron stars. In more massive stars ($\geq 20M_{sun}$), a neutron star forms by the collapse of the Fe core. Gravitational binding energy is released as neutrinos and they interact with heated material above the neutron star, generating a neutron-driven winds (e.g., Takahashi, Witti, & Janka 1994, Woosley et al. 1994, Wanajo et al. 2001). Neutron star mergers are also considered as a possible site for the r-process (Freiburghaus, Rosswog, & Thielemann 1999). However, neutron-star mergers seem to be ruled out as a dominant r-process source because their low rates of occurrence would lead to r-process enrichment that is not consistent with observation at very low metallicities (Argast et al. 2004).

As detailed above, the r-process and the p-process operate in different physical conditions and the time frames. Yet in diamond separates from meteorites, Xe-L and Xe-H are always correlated: the higher the former, the higher the latter. This correlation has been observed in all diamond separates from any types of meteorites. A possible explanation is that there might be diamond particles with only Xe-L and those with only Xe-H, however, they are completely mixed in meteorites. Thus, the correlation we see does not imply that Xe-L and Xe-H are in the same diamond particles, but that two kinds of diamond particles are completely homogenized in meteorites.

**3.2. r-Process $^{134}$Xe/$^{136}$Xe and p-process $^{124}$Xe/$^{126}$Xe ratios inferred from Xe-HL**

Another, more intriguing problem is that the p-process and the r-process isotopic ratios inferred from Xe-HL are different from those inferred from the solar system abundance (Ott 1996) and from model calculations of supernovae (Rayet et al. 1995). Xe-HL must be a mixture of at least three components, two of which are the p-process Xe and the r-process Xe. Another component is required because Xe-HL contains $^{130}$Xe that is not produced by either the p-process or the r-process. Assuming that the third component is solar, the r-process $^{134}$Xe/$^{136}$Xe and the p-process $^{124}$Xe/$^{126}$Xe in diamond inferred from Xe-HL are calculated to be 0.699 and 2.205, respectively (Ott 1996).

Those ratios are very different from the ratios derived from the solar system abundance. Although $^{134}$Xe is dominantly made by the r-process, a small portion of $^{134}$Xe is produced by the s-process, especially in low-metallicity stars (Pignatari et al. 2004). Thus, the solar system $^{134}$Xe and $^{136}$Xe abundances are corrected for the s-process Xe, subtracting the s-process Xe inferred from the noble gas analysis of the Murchison SiC separates (Lewis, Amari, & Anders 1994) from solar Xe (Pepin, Becker, & Rider 1995). The r-process $^{134}$Xe/$^{136}$Xe inferred from the solar system abundance is 1.207, whereas the p-process $^{124}$Xe/$^{126}$Xe from the solar system abundance is 1.157 (Ott 1996). Rayet et al. (1995) calculated p-process yields from the supernovae explosion of solar metallicity with masses of 13, 15, 20 and 25 $M_{sun}$. The average $^{124}$Xe/$^{126}$Xe for the 4 cases is 1.050 and is very close to the solar system $^{124}$Xe/$^{126}$Xe (Table 1).

Table 1. p- and r-process Xe isotopic ratios inferred from Xe-HL and the solar system abundance as well as p-process ratio predicted from models

|  | $^{124}$Xe/$^{126}$Xe | $^{134}$Xe/$^{136}$Xe | Reference |
|---|---|---|---|
| p- and r-processes in diamond | 2.205 | 0.699 | Ott (1996) |
| Solar | 1.157 | 1.207* | Pepin et al. (1995) |
| Supernova models | 1.050** |  | Rayet et al. (1995) |

\* Corrected for a small contribution from the s-process. See the text and Ott (1996).
\*\* Average of 13, 15, 20 and 25$M_{sun}$ stars.

In order to explain the difference in the r-process $^{134}$Xe/$^{136}$Xe ratios in the diamond and in the solar system, Ott (1996) proposed "a rapid separation scenario". Radiogenic precursors of $^{136}$Xe, $^{136}$Te ($T_{1/2}$ = 17.5 s) and $^{136}$I ($T_{1/2}$ = 1.39 min), have shorter half-lives than those of $^{134}$Xe, $^{134}$Te ($T_{1/2}$ = 42 min) and $^{134}$I ($T_{1/2}$ = 53 min). Therefore, if Xe is separated from Te and I before $^{134}$Te and $^{134}$I completely decay to $^{134}$Xe, the low $^{134}$Xe/$^{136}$Xe inferred from Xe-H can be explained. Ott (1996) estimated that the separation should take place 7900 seconds after the end of the r-process. As the $^{134}$Xe/$^{136}$Xe rapidly changes with time [see Fig. 2 in Ott (1996)], the timing of the separation has a very narrow window of tolerance. Ott (1996) explained the p-process $^{124}$Xe/$^{126}$Xe inferred from Xe-L in the same manner, using the difference in half-lives of $^{124}$Ba ($T_{1/2}$ = 12 min) and $^{124}$Cs ($T_{1/2}$ = 30 s), and those of $^{126}$Ba ($T_{1/2}$ = 1.65 h) and $^{126}$Cs ($T_{1/2}$ = 1.64 min). He argued that the separation of Xe, and Ba and Cs on a timescale of hours would produce a $^{124}$Xe/$^{126}$Xe ratio higher than solar.

There are a few difficulties in this scenario: first, the separation requires a very fine tuning of the timing as the $^{134}$Xe/$^{136}$Xe ratio changes so quickly. In addition, it is not yet clear how the separation of elements could be achieved less than 2 hours after the r-process occurred. However, this is among the first effort to explain the nucleosynthetic pattern of Xe in presolar diamond.

Another scenario to reproduce Xe-H is to mix a product of neutron burst, or mini r-process, and the solar component (Maas et al. 2001; Ott 2002). Heymann and Dziczkaniec (1979, 1980) explored a possibility that Xe-H is produced in the carbon burning shell, where temperatures are too high and neutron densities and doses are too small for the classical r-process. Later, Howard et al. (1992) refined neutron burst models to explain Xe-H. Meyer et al. (2000) revisited this mechanism to reproduce the Mo isotopic ratios of SiC X grains of supernova origin (Pellin et al.

1999). Neutron burst takes place in the outer He-rich region heated at T = $10^9$K when the supernova shock wave traverses it. Neutrons produced by $^{22}$Ne($\alpha$, n)$^{25}$Mg are captured by nuclei that were originally of a solar abundance and later had been exposed to a weak neutron fluence ($\tau$=0.002mb), to mimic the weak s-processing in the pre-supernova phase. If the material that had experienced the neutron burst is mixed with solar material with a ratio of ~20:80, the $^{134}$Xe/$^{136}$Xe ratio in the diamond can be reproduced (Ott 2002). In this case, however, Xe-L is left unexplained.

## 4. Isotopic analysis of trace elements in diamond separates

Richter et al. (1998) and Maas et al. (2001) analyzed Te isotopes in a diamond separate from the Allende meteorite. Although Richter et al. (1998) observed isotopically anomalous Te in the first steps of their analysis [as shown in Figure 1 in Richter et al. (1998)], the total Te in the diamond separate was isotopically close to normal (U. Ott 2006, private communication). Maas et al. (2001) found excesses in r-process only isotopes $^{128}$Te and $^{130}$Te, but they are very small: $(^{128}$Te/$^{124}$Te$)_{diamond}/(^{128}$Te/$^{124}$Te$)_{solar}$ = 1.0040 ± 0.0015 and $(^{130}$Te/$^{124}$Te$)_{diamond}/(^{130}$Te/$^{124}$Te$)_{solar}$ = 1.0093 ± 0.0028. It is puzzling why Te, a neighboring element of Xe, shows such a small anomaly, whereas the Xe isotopic anomaly is so pronounced. Small isotopic anomalies of Te may be due to impurities in the diamond separates. Trace element concentrations of diamond is extremely low (Lewis et al. 1991) and even a very small amount of impurities would compromise analysis. In fact, Ir-rich nuggets are known to be present in an Allende diamond separate (Lewis et al. 1991).

## 5. Concluding remarks

More than three decades have elapsed since the discovery of Xe-HL and two decades have passed since the isolation of presolar diamond from meteorites. Yet diamond remains enigmatic. The presence of Xe-HL points toward supernova origin of the Xe and (at least part of) diamond. However, the p-process ratio and the r-process ratio inferred from Xe-HL do not agree with those from the solar system abundance and the p-process ratio does not agree with p-process isotopic ratios predicted from stellar models.

Isotopic anomalies of heavy elements in the vicinity of Xe are bewilderingly small compared with the anomaly in the Xe. It may be attributed to a very low concentration of the heavy elements and contamination from small amounts of other minerals in the diamond separates. In the future, isotopic analysis of heavy elements in high-purity diamond separates by resonant ionization mass spectrometry (RIMS), where elements of interest are selectively ionized and detected by a time-of-flight mass spectrometer (e.g., Nicolussi et al. 1998; Savina et al. 2003) may be able to better answer a few outstanding questions.

**Acknowledgements**

This work is supported by NASA grant NNX08AG56G.